\documentclass[showpacs,aip,jmp,nofootinbib,twocolumn,12pt]{revtex4-1}%
\usepackage{amsmath}
\usepackage{amssymb}
\usepackage{dcolumn}
\usepackage{graphicx}
\usepackage{dcolumn}
\usepackage{bm}%
\usepackage{amsfonts}
\providecommand{\U}[1]{\protect\rule{.1in}{.1in}}
\begin{document}
\title{Killing-Yano tensors in spaces admitting a hypersurface orthogonal Killing vector}
\author{David Garfinkle}
\affiliation{Physics Department, Oakland University, Rochester, MI, 48309, USA}
\affiliation{Michigan Center for Theoretical 
Physics, Randall Laboratory of Physics, University of Michigan, Ann Arbor, MI 48109-1120, USA}
\author{E.N. Glass}
\affiliation{Michigan Center for Theoretical 
Physics, Randall Laboratory of Physics, University of Michigan, Ann Arbor, MI 48109-1120, USA}
%\date{\today}

\begin{abstract}
Methods are presented for finding Killing-Yano tensors, conformal Killing-Yano
tensors, and conformal Killing vectors in spacetimes with a hypersurface
orthogonal Killing vector. These methods are similar to a method developed by
the authors for finding Killing tensors. In all cases one decomposes both the
tensor and the equation it satisfies into pieces along the Killing vector and
pieces orthogonal to the Killing vector. Solving the separate equations that
result from this decomposition requires less computing than integrating the
original equation. In each case, examples are given to illustrate the method.

\end{abstract}

\pacs{04.20.Cv, 04.20.Jb}
\maketitle

\section{Introduction}

Recently Garfinkle and Glass \cite{GG10} presented a method for finding Killing tensors in 
spaces with a hypersurface orthogonal Killing vector.  The method involves a 3+1 (or more generally
($n$-1)+1) decomposition of the Killing tensor equation using the foliation orthogonal to the Killing
vector. The approach of \cite{GG10} has been
considered by Mirshekari and Will \cite{MW10} in showing that the Bach-Weyl
metric does not admit a non-trivial Killing tensor. Since the Killing tensor equation is one of a class
of similar tensor equations (Killing vector, conformal Killing vector, Killing-Yano, conformal Killing-Yano, etc.) 
it is natural to ask whether the approach of \cite{GG10} could be used on any of these other equations.
In fact, the use of 3+1 decomposition to study the equations for a Killing vector has a long history in general 
relativity begining with the work of Moncrief\cite{Mon75} and Coll\cite{Coll77} and continued {\it e.g.} by Beig and
Chru\'{s}ciel \cite{BC97}. More recently G\'{o}mez-Lobo
and Valiente-Kroon \cite{GLK08} considered this 3+1 decomposition in spinor formalism,
and have also studied Killing spinor initial data sets. The main difference between these earlier works and the 
method of \cite{GG10} is the assumption of a hypersurface orthogonal Killing vector.  This assumption greatly 
restricts the cases to which the method applies; however it also provides a great simplification to the equations
and thus makes them more tractable.  A similar approach due to  Bona and Coll
\cite{BC91} treats the conformal Killing
equation in static spacetimes, but then adds the further condition that 
the conformal Killing field is Lie derived by the static Killing field.  

This paper generalizes the technique of \cite{GG10} by producing analogous methods for the
Killing-Yano, conformal Killing, and conformal Killing-Yano equations.  In each case the spacetime
is assumed to posess a hypersurface orthogonal Killing vector, and the equations are decomposed with
respect to the foliation orthogonal to the Killing vector.  As a simple illustration of these techniques,
we find the Killing-Yano tensors of the Bertotti-Robinson (BR) spacetime, and the
conformal Killing-Yano tensors of a particular cylindrical
vacuum metric due to Linet.\cite{Lin85} \newline\textbf{Notation}: Lower case Latin indices, $B^{a}$,
range over n-dimensions. Greek indices, $B^{\mu}$, range over n--1 dimensions.
For Killing vector $\xi^{a}$ an overdot will denote a Lie derivative, $\dot
{A}:=\mathcal{L}_{\xi}A.$

\section{The Killing-Yano tensor method}

The Killing-Yano (KY) equation for antisymmetric tensor $A_{ab}$ can be
written as
\begin{equation}
A_{a(b;c)}=0.\label{ky-eqn}%
\end{equation}
This generalizes Killing's equation to antisymmetric tensors. There are at
most 10 independent solutions of the KY equation on manifold $\mathcal{M}$.
The maximum of 10 occurs if, and only if, $\mathcal{M}$ has constant curvature.
There is an extensive literature covering KY tensors. In an early paper
Collinson \cite{Col76} discussed the relationship between Killing vectors and
KY tensors. He pointed out that all type D vacuum solutions which admit a
Killing tensor also admit a KY tensor. Two works by Dietz and R\"{u}diger
\cite{DR81,DR82} discuss the character of spacetimes admitting KY tensors.
More recently, Ferrando and S\'{a}ez \cite{FS03} gave Rainich conditions for
systems to admit KY tensors. Hall \cite{Hal87} studied the existence of KY
tensors in General Relativity, and Ibohal \cite{Ibo97} has used the
Newman-Penrose formalism to integrate the KY equations and has found a number
of spacetimes which contain KY tensors, including FRW, Kerr-Newman, and
Bertotti-Robinson \cite{Ber59}. Taxiarchis \cite{Tax85} has proved that the
only spacetimes which admit KY tensors have Petrov type D, N, or O.

Suppose that a spacetime has a hypersurface orthogonal Killing vector $\xi
^{a}$. Define $V$ such that
\begin{equation}
{\xi^{a}}{\xi_{a}}=\epsilon{V^{2}}%
\end{equation}
where $\epsilon=\pm1$. Then the metric in directions orthogonal to $\xi^{a}$
is given by
\begin{equation}
{h_{ab}}={g_{ab}}-\epsilon{V^{-2}}{\xi_{a}}{\xi_{b}}%
\end{equation}
One can use ${h^{a}}_{b}$ as a projection operator to project any tensor in
directions orthogonal to $\xi^{a}$. In particular, the KY tensor can be
decomposed as
\begin{equation}
{A_{ab}}=2V^{-1}S{_{[a}}{\xi_{b]}}+Q{_{ab}}\label{decomposition}%
\end{equation}
where $S_{a}$ and antisymmetric $Q_{ab}$ are orthogonal to $\xi^{a}$.

Projecting the KY equation using all combinations of ${h^{a}}_{b}$ and
$\xi^{a}$ yields the following
\begin{align}
{D_{a}Q}{_{bc}+D}_{b}Q_{ac}  & =0,\label{ky1}\\
{D_{(a}}{S_{b)}}  &  =0,\label{ky2}\\
{\mathcal{L}_{\xi}}{Q_{ab}}  &  = \epsilon{V^{3}}{D_{[a}V}^{-2}S_{b]}%
,\label{ky3}\\
{\mathcal{L}_{\xi}S}{_{a}}  &  = -{Q}_{ab}{D}^{b}V{.}\label{ky4}%
\end{align}
Here $\mathcal{L}_{\xi}$ denotes the Lie derivative with respect to Killing
vector $\xi^{a}$ and $D_{a}$ denotes the derivative operator on the space
orthogonal to $\xi^{a}$.

The first two equations say that $Q_{ab}$ and $S^{a}$ are respectively a
Killing-Yano tensor and a Killing vector on the space orthogonal to $\xi^{a}$.
The last two equations are additional conditions that these tensors must
satisfy. These last two equations are most easily implemented in a coordinate
system adapted to the Killing vector. Choose a coordinate system ($y,{x^{\mu}%
}$) such that $x^{\mu}$ are coordinates on the surface orthogonal to the
Killing vector and $\mathcal{L}_{\xi}$ is simply a partial derivative with
respect to $y$. Use $\partial_{\mu}$ or a comma to denote a derivative with
respect to the $x^{\mu}$ coordinates. The Latin indices in this section are
n-dimensional, and the method below projects objects and equations down to n-1
dimensions with Greek indices.

Equations (\ref{ky3}-\ref{ky4}) become%
\begin{align}
\dot{Q}_{\mu\nu}  &  = \epsilon V\partial_{\lbrack\mu}S_{\nu]}+2\epsilon
S_{[\mu}\partial_{\nu]}V\label{ky5}\\
\dot{S}_{\mu}  &  = -Q_{\mu\alpha}h^{\alpha\nu}\partial_{\nu}V.\label{ky6}%
\end{align}

Thus the method for finding Killing-Yano tensors on the n-dimensional space
consists of two steps: \newline(1) find all Killing-Yano tensors and all
Killing vectors on the $n-1$ dimensional space \newline(2) subject those
Killing-Yano tensors and Killing vectors to the conditions of Eq.(\ref{ky5})
and\hspace{0.1in}Eq.(\ref{ky6})

\section{Killing-Yano tensors of the Bertotti-Robinson metric}

The BR spacetime (up to an overall scale) has line element
\begin{align}
d{s^{2}}=(\frac{1}{r^{2}})(-d{t^{2}}+dr^{2}\nonumber\\
+r^{2}d\vartheta^{2}+r^{2}\sin^{2}\vartheta d{\varphi}^{2})\label{br-met-1}%
\end{align}
This spacetime is the direct product of the 2-sphere and 2-dimensional anti
de-Sitter spacetime, i.e. $S^{2}\otimes AdS_{2}$. Defining coordinate $w:=-\ln
r$ allows the BR line element to be transformed to the form
\begin{equation}
d{s^{2}}=-{e^{2w}}d{t^{2}}+d{w^{2}}+d\vartheta^{2}+\sin^{2}\vartheta
d{\varphi}^{2}.\label{br-met-2}%
\end{equation}
For the convenience of the reader in following this section, additional properties of the 
BR spacetime are collected in Appendix A.

The method of the previous section will be used to work out the KY tensors of
the BR metric. First the KY tensors of the 2-dimensional $w\vartheta$ surface
will be found, then these will be used to find the KY tensors of the
3-dimensional $w\vartheta\varphi$ surface, and finally find the KY tensors of
4-dimensional BR spacetime.

${c_{1}},\,{c_{2}}$ etc. will denote constants, and ${k_{1}},\,{k_{2}}$ etc.
will denote quantities that depend only on the coordinate associated with the
Killing vector.

\subsection*{$w\vartheta$ and $w\vartheta\varphi$ surfaces}

The 2-dimensional $w\vartheta$ space has line element
\begin{equation}
d{s^{2}}=d{w^{2}}+d\vartheta^{2}.
\end{equation}
For any 2-dimensional space, the unique solution (up to an overall scale) of
Eq.(\ref{ky1}) is the volume element. Since the $w\vartheta$ space is just
ordinary 2-dimensional Euclidean space, it has the three Killing vectors of
that space. Thus we have
\begin{align}
{Q_{\mu\nu}}  & ={k_{1}2}\partial_{\lbrack\mu}w\ {\partial}_{\nu]}\vartheta\\
{S_{\mu}}  & ={k_{2}}{\partial_{\mu}}w+{k_{3}}{\partial_{\mu}}\vartheta
\nonumber\\
& +{k_{4}}(\vartheta{\partial_{\mu}}w-w{\partial_{\mu}}\vartheta)
\end{align}
Using the $\varphi$ Killing vector of metric (\ref{br-met-2}) and recalling
that the Killing vector norm is $\epsilon V^{2}$, yields $V=\sin\vartheta$ and
$\epsilon=1$. Imposing Eq.(\ref{ky6}) we find
\begin{align}
{{\dot{k}}_{2}}{\partial_{\mu}}w+{{\dot{k}}_{3}}{\partial_{\mu}}%
\vartheta+{{\dot{k}}_{4}}(\vartheta{\partial_{\mu}}w-w{\partial_{\mu}%
}\vartheta)\nonumber\\
=-{k_{1}}\cos\vartheta({\partial_{\mu}}w)
\end{align}
It then follows that the quantities ${k_{1}},\,{{\dot{k}}_{2}},\,{{\dot{k}%
}_{3}}$ and ${{\dot{k}}_{4}}$ all vanish. Thus we have $Q_{\mu\nu}=0$ and
\begin{equation}
{S_{\mu}}={c_{2}}{\partial_{\mu}}w+{c_{3}}{\partial_{\mu}}\vartheta+{c_{4}%
}(\vartheta{\partial_{\mu}}w-w{\partial_{\mu}}\vartheta)
\end{equation}
Now, using Eq.(\ref{ky5}) we find
\begin{align}
0=({c_{4}}\sin\vartheta+{c_{2}}\cos\vartheta+{c_{4}}\vartheta\cos
\vartheta)\nonumber\\
\times{(}\partial_{\mu}w\ {\partial}_{\nu}\vartheta-\partial_{\nu}%
w\ \partial_{\mu}\vartheta).
\end{align}
This implies that $c_{2}$ and $c_{4}$ vanish. It follows that ${S_{\mu}%
}={c_{3}}{\partial_{\mu}}\vartheta$. Use of Eq.(\ref{decomposition}) results
in
\begin{equation}
{A_{\mu\nu}}={c_{3}}\sin\vartheta\ 2{\partial_{\lbrack\mu}}\vartheta
\ {\partial_{\nu]}}\varphi
\end{equation}

\subsection*{the $w\vartheta\varphi$ surface and the BR spacetime}

The 3-dimensional $w\vartheta\varphi$ space has line element
\begin{equation}
d{s^{2}}=d{w^{2}}+d\vartheta^{2}+{\sin^{2}}\vartheta d{\varphi^{2}}%
\end{equation}
Here the Killing vectors are ${(\partial/\partial w)}^{a}$ and the three
Killing vectors of the 2-sphere, which will be denoted by ${\xi^{1a}},$
${\xi^{2a},}$ $\xi^{3a}$. We therefore have
\begin{equation}
{S_{\mu}}={k_{2}}{\partial_{\mu}}w+{k_{3}}{\xi_{\mu}^{1}}+{k_{4}}{\xi_{\mu
}^{2}}+{k_{5}}{\xi_{\mu}^{3}}%
\end{equation}
From the results of the previous subsection it follows that
\begin{equation}
{Q_{\mu\nu}}={k_{1}}\sin\vartheta\ 2{\partial_{\lbrack\mu}}\vartheta
\ {\partial_{\nu]}}\varphi
\end{equation}
The $t$ Killing vector of metric (\ref{br-met-2}) provides $\epsilon=-1$ and
$V={e^{w}}$. Using Eq.(\ref{ky6}) we have
\begin{equation}
0={{\dot{k}}_{2}}{\partial_{\mu}}w+{{\dot{k}}_{3}}{\xi_{\mu}^{1}}+{{\dot{k}%
}_{4}}{\xi_{\mu}^{2}}+{{\dot{k}}_{5}}{\xi_{\mu}^{3}}%
\end{equation}
Since the terms on the right hand side are linearly independent, the
coefficient of each term vanishes. Thus ${{\dot{k}}_{2}}={{\dot{k}}_{3}%
}={{\dot{k}}_{4}}={{\dot{k}}_{5}}=0$. Therefore one has ${k_{2}}={c_{2}%
},\,{k_{3}}={c_{3}},\,{k_{4}}={c_{4}},$ ${k_{5}}={c_{5}}$. It then follows
that $S_{\mu}$ takes the form
\begin{equation}
{S_{\mu}}={c_{2}}{\partial_{\mu}}w+\ell_{\mu}%
\end{equation}
where $\ell_{\mu}$ is the sum of 2-sphere Killing vectors, defined as%
\begin{equation}
{\ell_{\mu}:}={c_{3}}{\xi_{\mu}^{1}}+{c_{4}}{\xi_{\mu}^{2}}+{c_{5}}{\xi_{\mu
}^{3}.}%
\end{equation}
Upon using Eq.(\ref{ky6}) we find
\begin{align}
{{\dot{k}}_{1}}\sin\vartheta\ 2{\partial_{\lbrack\mu}}\vartheta\ {\partial
_{\nu]}}\varphi\nonumber\\
=-{e^{w}}{\partial_{\lbrack\mu}}{\ell_{\nu]}}-{e^{w}\ 2}{\ell_{\lbrack\mu}%
}{\partial_{\nu]}}w
\end{align}
The last term on the right hand side is linearly independent of both the first
curl term on the right hand side and the term on the left hand side. It then
follows that this term must vanish. Therefore ${\ell_{\mu}}=0$, and the entire
right hand side of this equation vanishes. The left hand side must therefore
also vanish, and so ${{\dot{k}}_{1}}=0$. Thus ${k_{1}}={c_{1}} $. Finally we
have
\begin{align}
{Q_{\mu\nu}}  &  ={c_{1}}\sin\vartheta\ 2{\partial_{\lbrack\mu}}%
\vartheta\ {\partial_{\nu]}}\varphi,\\
{S_{\mu}}  &  ={c_{2}}{\partial_{\mu}}w.
\end{align}
Applying this result in Eq.(\ref{decomposition}) we find that the general
Killing-Yano tensor of the BR spacetime is
\begin{align}
{A_{\mu\nu}}=({c_{1}}\sin\vartheta)\ 2{\partial_{\lbrack\mu}}\vartheta
\ {\partial_{\nu]}}\varphi\nonumber\\
+({c_{2}}{e^{w})\ 2}{\partial_{\lbrack\mu}}t\ {\partial_{\nu]}}w
\end{align}
Since the BR spacetime is the direct product $S^{2}\otimes AdS_{2}$, this
result has a simple geometrical interpretation. The Killing-Yano tensors of
the BR spacetime are the volume elements of $S^{2}$ and $AdS_{2}$.

\section{Conformal Killing-Yano Tensors}

The tensor version of the conformally covariant generalization of the KY
equation, the CKY equation, was discovered by Tachibana \cite{Tac69}. It can
be written in the form
\begin{align}
& {\nabla_{a}}{A_{bc}}+{\nabla_{b}}{A_{ac}}\nonumber\\
& =2{W_{c}}{g_{ab}}-{W_{a}}{g_{bc}}-{W_{b}}{g_{ac}}\label{cky}%
\end{align}
for some $W_{a}$. ${A_{bc}}$ is given in Eq.(\ref{decomposition}). It follows
from Eq.(\ref{cky}) that
\begin{equation}
{W_{a}}={\frac{1}{n-1}}{\nabla^{b}}{A_{ba}}\,.
\end{equation}
In \cite{Tac69} Tachibana showed that in a Ricci-flat space, for $A_{ab}$ a
CKY bivector satisfying Eq.(\ref{cky}), $(1/3)\nabla^{b}A_{ab}$ is a Killing
vector. It is well known \cite{GK99} that the Kerr metric admits a CKY
bivector, and indeed all type D vacuum solutions and their charged counterparts
have a CKY bivector.\cite{FS07}

In a manner just like the Killing-Yano case, we can decompose $A_{ab}$ as in
Eq.(\ref{decomposition}). Similarly, $W_{a}$ can be decomposed as
\begin{equation}
{W_{a}}=\gamma{V^{-1}}{\xi_{a}}+{X_{a}}%
\end{equation}
where $X_{a}$ is orthogonal to $\xi^{a}$. Taking all projections of
Eq.(\ref{cky}) we find the following:
\begin{align}
{D_{a}Q}{_{bc}+D}_{b}Q_{ac}  & =\nonumber\\
2{X_{c}}{h_{ab}}-{X_{a}}{h_{bc}} & -{X_{b}}{h_{ac}},\label{cky1}\\
{D_{(a}}{S_{b)}}  &  =\gamma{h_{ab}},\label{cky2}\\
{\mathcal{L}_{\xi}}{Q_{ab}}  & =\epsilon{V^{3}}{D_{[a}V}^{-2}S_{b]}%
,\label{cky3}\\
{\mathcal{L}_{\xi}S}{_{a}}  & =-{Q}_{ab}{D}^{b}V-V{X_{a}}{.}\label{cky4}%
\end{align}
As in the Killing-Yano case, the first two equations have a simple geometrical
interpretation. On the $n-1$ dimensional subspace orthogonal to the Killing
vector $Q_{ab}$ is a CKY tensor and $S^{a}$ is a conformal Killing vector. The
last two equations provide conditions that those tensors must satisfy.

We now specialize to the case where $n=4$ and the spacetime has a Lorentz
signature. A vector $T^{a}$ exists such that%
\begin{equation}
Q_{ab}=\epsilon_{abc}T^{c}\label{Q-T-eqn}%
\end{equation}
where $\epsilon_{abc}$ is the volume element of the 3-dimensional space
orthogonal to the Killing vector. Then equations (\ref{cky1}), (\ref{cky3}),
and (\ref{cky4}) become
\begin{align}
D_{(a}T_{b)}  & = \psi h_{ab}\label{D-T-psi}\\
{\mathcal{L}_{\xi}T}^{a}  & = -\frac{1}{2}V^{3}\epsilon^{abc}D_{b}(V^{-2}%
S_{c})\label{Lie-T}\\
{\mathcal{L}_{\xi}S}^{a}  & = \frac{1}{2}V^{3}\epsilon^{abc}D_{b}(V^{-2}%
T_{c})\label{Lie-S}%
\end{align}
Thus ${T}^{a}$ is a conformal Killing vector of the 3-dimensional space.  
The additional conditions that ${T}^{a}$ and ${S}^{a}$ must
satisfy are given by equations (\ref{Lie-T}) and (\ref{Lie-S}) respectively.
In the adapted coordinate system these additional conditions take the form%
\begin{align}
\dot{T}^{\mu}  &  =-\frac{1}{2}V^{3}\epsilon^{\mu\nu\alpha}\partial_{\nu
}(V^{-3}S_{\alpha})\label{T-dot}\\
\dot{S}^{\mu}  &  =\frac{1}{2}V^{3}\epsilon^{\mu\nu\alpha}\partial_{\nu
}(V^{-3}T_{\alpha})\label{S-dot}%
\end{align}
Thus to find the CKY tensors of the 4-dimensional spacetime, one does the
following: \newline(1) find all the conformal Killing fields ${S}^{a}$ and
${T}^{a}$ of the 3-dimensional surface orthogonal to the Killing vector.
\newline(2) subject those conformal Killing fields to the conditions of
equations (\ref{T-dot}) and (\ref{S-dot}). \newline(3) use Eq.(\ref{Q-T-eqn})
to find $Q_{ab}$ and then Eq.(\ref{decomposition}) to find $A_{ab}$.

\section{Conformal Killing Vectors}
Since, as shown in the previous section, one step in finding CKY tensors involves
finding conformal Killing vectors, we now apply the general method of this paper 
to finding conformal Killing vectors.  Recall that a
conformal Killing vector $K^{a}$ on an n-dimensional space is one for which%
\begin{equation}
\nabla_{a}K_{b}+\nabla_{b}K_{a}=\frac{2}{n}(\nabla_{c}K^{c})g_{ab}%
.\label{cfk-eqn}%
\end{equation}
In a manner similar to the Killing-Yano case, we can decompose $K_{a}$ as%
\begin{equation}
K_{a}=A\xi_{a}+B_{a}\label{ckvdecomp}%
\end{equation}
where $B_{a}$ is orthogonal to $\xi^{a}$. Taking all projections of
Eq.(\ref{cfk-eqn}) it follows that%
\begin{align}
D_{a}B_{b}+D_{b}B_{a}  &  =\frac{2}{n-1}(D_{c}B^{c})h_{ab}\label{DB-eqn}\\
{\mathcal{L}_{\xi}A\ }  &  {=\ }\frac{V^{n-1}}{n-1}D_{a}(V^{1-n}%
B^{a})\label{Lie-A}\\
D_{a}A  &  =-\epsilon V^{-2}{\mathcal{L}_{\xi}B}_{a}\label{DA-eqn}%
\end{align}
As in the Killing-Yano case, Eq.(\ref{DB-eqn}) has a simple geometrical
interpretation. On the $n-1$ dimensional subspace orthogonal to Killing vector
$\xi^{a}$, $B^{a}$ is a conformal Killing vector. However, this conformal
Killing vector is also subject to additional conditions, the integrability
conditions for $A$. Taking the curl of Eq.(\ref{DA-eqn}) we obtain%
\begin{equation}
D_{[a}(V^{-2}{\mathcal{L}_{\xi}B}_{b]})=0.
\end{equation}
Subtracting $D_{a}$ of Eq.(\ref{Lie-A}) from ${\mathcal{L}_{\xi}}$ of
Eq.(\ref{DA-eqn}) provides%
\begin{align}
& {\mathcal{L}_{\xi}}{\mathcal{L}_{\xi}}{B_{a}}\nonumber\\
& +\frac{\epsilon V^{2}}{n-1}D_{a}[V^{n-1}D_{b}(V^{1-n}B^{b})]=0.
\end{align}
In the adapted coordinate system, the additional conditions for $B^{a}$ become%
\begin{align}
& \partial_{\lbrack\mu}(V^{-2}\dot{B}_{\nu]}) =0\label{curl-Bdot}\\
\ddot{B}_{\mu} & +\frac{\epsilon V^{2}}{n-1}\partial_{\mu}\left[
\frac{V^{n-1}}{\sqrt{h}}\partial_{\nu}(\sqrt{h}V^{1-n}B^{\nu})\right]
\nonumber\\
& =0.\label{B-dot-dot}%
\end{align}
The equations for $A$ are%
\begin{align}
\dot{A}  &  =\frac{V^{n-1}}{(n-1)\sqrt{h}}\partial_{\nu}(\sqrt{h}V^{1-n}%
B^{\nu})\label{A-dot}\\
\partial_{\mu}A  &  =-\epsilon V^{-2}\dot{B}_{\mu}\label{grad-A}%
\end{align}
Thus the method for finding conformal Killing vectors on the n-dimensional
space consists of three steps: \newline(1) find all conformal Killing vectors
on the $n-1$ dimensional space \newline(2) subject those conformal Killing
vectors to the conditions of Eq.(\ref{curl-Bdot}) and\hspace{0.1in}%
Eq.(\ref{B-dot-dot}) \newline(3) solve Eq.(\ref{A-dot}) and Eq.(\ref{grad-A})
for $A$.

\section{CKY Tensor of Linet's Vacuum Metric}

The Petrov type D cylindrical vacuum line element found by Linet \cite{Lin85}
is written as%
\begin{equation}
ds^{2}=r^{4}(-dt^{2}+dr^{2}+dz^{2})+r^{-2}d\varphi^{2}.
\end{equation}
This static metric has Killing vectors $\partial_{t},$ $\partial_{z},$
$\partial_{\varphi}$. We will use the method of the previous two sections to find the CKY
tensors of this spacetime.

\subsection*{$tr$ and $tr\varphi$ surfaces}

We begin by finding all the conformal Killing fields of the $tr$ surface and
then using those to find all the conformal Killing fields of the $tr\varphi$
surface. The 2-dimensional $tr$ space has line element%
\begin{equation}
ds^{2}=r^{4}(-dt^{2}+dr^{2}).
\end{equation}
Like all 2-dimensional metrics, this metric is conformally flat and the
conformal Killing fields are therefore those of the underlying flat spacetime.
For our purposes, it will be convenient to use null coordinates $u=t-r$ and
$v=t+r$. The line element then becomes%
\begin{equation}
ds^{2}=-r^{4}dudv\label{du-dv-met}%
\end{equation}
where $r=(v-u)/2$. It follows from metric (\ref{du-dv-met}) that the conformal
Killing field takes the form%
\begin{equation}
B^{a}=\alpha(\partial_{u})^{a}+\beta(\partial_{v})^{a}\label{conf-B-vec}%
\end{equation}
where $\alpha$ is independent of $v$, and $\beta$ is independent of $u$. We
now use this conformal Killing field to work out the general conformal Killing
field of the $tr\varphi$ surface. We have $g_{\varphi\varphi}=r^{-2}$,
therefore $\epsilon=1$ and $V=r^{-1}$. It then follows that%
\begin{align}
& {\partial_{\lbrack\mu}}(V^{-2}\dot{B}_{\nu]})= {\textstyle \frac{1}{2}}%
r^{6}[-\partial_{u}\dot{\alpha}+\partial_{v}\dot{\beta}\nonumber\\
& +3r^{-1}(\dot{\alpha}+\dot{\beta})]\ \partial_{\lbrack\mu}u\ \partial_{\nu
]}v
\end{align}
and therefore from Eq.(\ref{curl-Bdot})%
\begin{equation}
-\partial_{u}\dot{\alpha}+\partial_{v}\dot{\beta}+3r^{-1}(\dot{\alpha}%
+\dot{\beta})=0.\label{a-dot-b-dot-1}%
\end{equation}
Taking $\partial_{u}\partial_{v}$ of this equation, and using the fact that
$\alpha$ is independent of $v$ and $\beta$ is independent of $u$, we find%
\begin{equation}
-\partial_{u}\dot{\alpha}+\partial_{v}\dot{\beta}-r^{-1}(\dot{\alpha}%
+\dot{\beta})=0.\label{a-dot-b-dot-2}%
\end{equation}
Subtracting Eq.(\ref{a-dot-b-dot-2}) from Eq.(\ref{a-dot-b-dot-1}) yields
\begin{equation}
\dot{\alpha}+\dot{\beta}=0.\label{a-dot-b-dot-3}%
\end{equation}
However, since $\alpha$ is independent of $v$ and $\beta$ is independent of
$u$, there exists a function $k_{1}(\varphi)$ such that $\dot{\alpha}=-k_{1}$
and $\dot{\beta}=k_{1}$. It then follows from Eq.(\ref{conf-B-vec}) that%
\begin{equation}
\dot{B}^{a}=-k_{1}(\partial_{u})^{a}+k_{1}(\partial_{v})^{a}=k_{1}%
(\partial_{r})^{a}.\label{Bdot}%
\end{equation}
Now taking $\partial_{\varphi}$ of Eq.(\ref{B-dot-dot}) for ${{\ddot{B}}_{\mu
}}$, and using Eq.(\ref{Bdot}) we find
\begin{align}
0  & = {{\dddot{B}}_{\mu}}+{\frac{\epsilon{V^{2}}}{n-1}}{\partial_{\mu}%
}\left[  {\frac{V^{n-1}}{\sqrt{h}}}{\partial_{\nu}}(\sqrt{h}{V^{1-n}}{{\dot
{B}}^{\nu}})\right] \nonumber\\
& =({r^{4}}{{\ddot{k}}_{1}}-3{r^{-4}}{k_{1}}){\partial_{\mu}}r
\end{align}
It then follows that ${k_{1}}=0$. Therefore ${{\dot{B}}^{\mu}}=0$ and so
$\alpha$ and $\beta$ are independent of $\varphi$. Thus $\alpha$ is a function
of $u$, and $\beta$ is a function of $v$. It then follows from
Eq.(\ref{B-dot-dot}) that there is a constant $c_{1}$ such that
\begin{align}
8{c_{1}} & ={\frac{V^{n-1}}{\sqrt{h}}}{\partial_{\nu}}(\sqrt{h}{V^{1-n}%
}{B^{\nu}})\nonumber\\
& ={\partial_{u}}\alpha+{\partial_{v}}\beta+3{r^{-1}}(\beta-\alpha
).\label{divB1}%
\end{align}
Differentiating Eq.(\ref{divB1}) by ${\partial_{u}}{\partial_{v}}$ yields
\begin{equation}
0={\partial_{u}}\alpha+{\partial_{v}}\beta-{r^{-1}}(\beta-\alpha
).\label{divB2}%
\end{equation}
Subtracting Eq.(\ref{divB2}) from Eq.(\ref{divB1}) provides
\begin{equation}
8{c_{1}}=4{r^{-1}}(\beta-\alpha)
\end{equation}
from which it follows that
\begin{equation}
\beta-\alpha=2{c_{1}}r={c_{1}}(v-u).
\end{equation}
But $\alpha$ depends only on $u$ and $\beta$ depends only on $v$ and so there
exists a constant $c_{2}$ such that
\begin{align}
\alpha &  ={c_{1}}u+{c_{2}}\\
\beta &  ={c_{1}}v+{c_{2}}%
\end{align}
We therefore have
\begin{align}
{B^{a}}  & = {c_{1}}\left[  u{{\left(  \partial_{u}\right)  }^{a}}+v{{\left(
\partial_{v}\right)  }^{a}} \right] \nonumber\\
& + {c_{2}} \left[  {{\left(  \partial_{u}\right)  }^{a}} +{{\left(
\partial_{v}\right)  }^{a}}\right] \nonumber\\
& ={c_{1}[}t{{\left(  \partial_{t}\right)  }^{a}}+r{{\left(  \partial
_{r}\right)  }^{a}]}+{c_{2}}{{\left(  \partial_{t}\right)  }^{a}}%
\end{align}
It then follows from Eq.(\ref{A-dot}) and Eq.(\ref{grad-A}) that ${\dot{A}%
}=4{c_{1}}$ and ${\partial_{\mu}}A=0$. We therefore have
\begin{equation}
A=4{c_{1}}\varphi+{c_{3}.}%
\end{equation}
Finally, using Eq.(\ref{ckvdecomp}) we find that the general conformal Killing
vector of the 3-dimensional $tr\varphi$ surface is
\begin{align}
& {K^{a}}={c_{1}}\left[  t{{\left(  \partial_{t}\right)  }^{a}}+r{{\left(
\partial_{r}\right)  }^{a}}+4\varphi{{\left(  \partial_{\varphi}\right)  }%
^{a}}\right] \nonumber\\
& +{c_{2}}{{\left(  \partial_{t}\right)  }^{a}}+{c_{3}}{{\left(
\partial_{\varphi}\right)  }^{a}}%
\end{align}

\subsection*{the $tr\varphi$ surface and the Linet spacetime}

The vector fields $T^{a}$ and $S^{a}$ are conformal Killing fields on the
$tr\varphi$ surface, with $z$ dependent coefficients and therefore take the
form
\begin{align}
{T^{a}}  & ={k_{1}}\left[  t{{\left(  \partial_{t}\right)  }^{a}}+r{{\left(
\partial_{r}\right)  }^{a}}+4\varphi{{\left(  \partial_{\varphi}\right)  }%
^{a}}\right] \nonumber\\
& + {k_{2}}{{\left(  \partial_{t}\right)  }^{a}}+{k_{3}}{{\left(
\partial_{\varphi}\right)  }^{a}}\\
{S^{a}}  &  ={k_{4}}\left[  t{{\left(  \partial_{t}\right)  }^{a}}+r{{\left(
\partial_{r}\right)  }^{a}}+4\varphi{{\left(  \partial_{\varphi}\right)  }%
^{a}}\right] \nonumber\\
& + {k_{5}}{{\left(  \partial_{t}\right)  }^{a}}+{k_{6}}{{\left(
\partial_{\varphi}\right)  }^{a}}%
\end{align}
Since ${g_{zz}}={r^{4}}$ it follows that $V={r^{2}}$ and $\epsilon=1$. We then
find
\begin{align}
& {\epsilon^{\mu\nu\lambda}}{\partial_{\nu}}({V^{-2}}{T_{\lambda}})
=\nonumber\\
& -6 {r^{-7}}{\epsilon^{\mu r\varphi}}(4{k_{1}}\varphi+{k_{3}})\\
& {\epsilon^{\mu\nu\lambda}}{\partial_{\nu}}({V^{-2}}{S_{\lambda}})
=\nonumber\\
& -6 {r^{-7}}{\epsilon^{\mu r\varphi}}(4{k_{4}}\varphi+{k_{6}})
\end{align}
From the $t$ component of Eq.(\ref{T-dot}) and Eq.(\ref{S-dot}) it follows
that
\begin{align}
{{\dot{k}}_{1}}t+{{\dot{k}}_{2}}  &  =-3{r^{-4}}(4{k_{4}}\varphi+{k_{6}}).\\
{{\dot{k}}_{4}}t+{{\dot{k}}_{5}}  &  =3{r^{-4}}(4{k_{1}}\varphi+{k_{3}}).
\end{align}
Therefore ${k_{1}},\,{k_{3}},\,{k_{4}},$ and $k_{6}$ vanish, and $k_{2}$ and
$k_{5}$ are constants. Thus the conformal Killing fields subject to
restrictions (\ref{T-dot}) and (\ref{S-dot}) take the form
\begin{align}
{T^{a}}  &  ={c_{1}}{{\left(  \partial_{t}\right)  }^{a}}\\
{S^{a}}  &  ={c_{2}}{{\left(  \partial_{t}\right)  }^{a}}%
\end{align}
Finally, using Eq.(\ref{Q-T-eqn}) and Eq.(\ref{decomposition}) we find that
the general CKY tensor of the spacetime is
\begin{equation}
{A_{\mu\nu}}=2{c_{1}}{r^{3}}{\partial_{\lbrack\mu}}r\ {\partial_{\nu]}}%
\varphi+2{c_{2}}{r^{6}}{\partial_{\lbrack\mu}}z\ {\partial_{\nu]}}t
\end{equation}

\section{Summary}

In this work a method is developed which decomposes the Killing-Yano tensor
into separate terms based on the surface geometry of metrics with a
hypersurface orthogonal Killing vector, and which thereby simplifies the
solution of the Killing-Yano equation. Using this method, we have shown that
the Bertotti-Robinson spacetime has a general KY tensor which is the sum of
volume bivectors. An enhancement of this method has also been applied to the
conformal Killing-Yano equation. The general CKY tensor has been constructed
for Linet's cylindrical vacuum metric.

\textbf{Acknowledgement }\ We would like to thank Jean Krisch for helpful
discussions. DG was supported by NSF Grants PHY-0855532 and PHY-1205202 to Oakland University.

\appendix{}

\section{Bertotti-Robinson}

The static Bertotti-Robinson (BR) metric is%
\begin{align}
& {g_{\mu\nu}^{\text{BR}}}dx^{\mu}dx^{\nu}=[(1+\lambda^{2}z^{2})dt^{2}%
-(1+\lambda^{2}z^{2})^{-1}dz^{2}]\nonumber\\
& - [(1-\lambda^{2}y^{2})dx^{2}+(1-\lambda^{2}y^{2})^{-1}dy^{2}%
].\label{br-met}%
\end{align}
$\lambda^{2}$ characterizes the electromagnetic energy density. Since the Weyl
tensor vanishes, the Petrov type is \textbf{0. }The BR spacetime has a
diagonal trace-free Ricci tensor (with rows and columns along $t,x,y,z$)%
\begin{equation}
\lbrack R_{\ \beta}^{\alpha}]^{\text{BR}}=\lambda^{2}\left[
\begin{array}
[c]{cccc}%
1 &  &  & \\
& -1 &  & \\
&  & -1 & \\
&  &  & 1
\end{array}
\right]  .\label{br-ricci}%
\end{equation}
The BR manifold is non-singular with Kretschmann scalar
\[
R_{\alpha\beta\mu\nu}R^{\alpha\beta\mu\nu}=8\lambda^{4}.
\]

The BR metric is spanned by the null tetrad
\begin{align}
l_{\alpha}dx^{\alpha}  & =(1/\sqrt{2})[(1+\lambda^{2}z^{2})^{1/2}dt\nonumber\\
& + (1+\lambda^{2}z^{2})^{-1/2}dz]\label{el}\\
n_{\alpha}dx^{\alpha}  & =(1/\sqrt{2})[(1+\lambda^{2}z^{2})^{1/2}dt\nonumber\\
& - (1+\lambda^{2}z^{2})^{-1/2}dz]\label{en}\\
m_{\alpha}dx^{\alpha}  & =(1/\sqrt{2})[(1-\lambda^{2}y^{2})^{1/2}dx\nonumber\\
& - i(1-\lambda^{2}y^{2})^{-1/2}dy]\label{em}\\
\bar{m}_{\alpha}dx^{\alpha}  & =(1/\sqrt{2})[(1-\lambda^{2}y^{2}%
)^{1/2}dx\nonumber\\
& + i(1+\lambda^{2}y^{2})^{-1/2}dy]\label{em-bar}%
\end{align}
Eight Newman-Penrose spin coefficients vanish, $\kappa=\sigma=\tilde{\lambda
}=\nu=\rho=\mu=\tau=\pi$. The remaining four are
\begin{align}
\epsilon &  =\gamma=-\frac{1}{2\sqrt{2}}\frac{\lambda^{2}z}{\sqrt
{1+\lambda^{2}z^{2}}}\label{ep-gam}\\
\alpha &  =\beta=-\frac{i}{2\sqrt{2}}\frac{\lambda^{2}y}{\sqrt{1-\lambda
^{2}y^{2}}}\label{alph-beta}%
\end{align}
The null vectors are all geodesic
\begin{align}
l_{\alpha;\beta}  &  =2\gamma l_{\alpha}l_{\beta}+2\gamma l_{\alpha}n_{\beta
}\label{dl}\\
n_{\alpha;\beta}  &  =-2\epsilon n_{\alpha}n_{\beta}-2\gamma n_{\alpha
}l_{\beta}\label{dn}\\
m_{\alpha;\beta}  &  =2\bar{\alpha}m_{\alpha}\bar{m}_{\beta}-2\alpha
m_{\alpha}m_{\beta}\label{dm}%
\end{align}
The BR manifold admits antisymmetric tensor $A_{\alpha\beta}$ as covariant
constant bivectors
\begin{align}
A_{\alpha\beta}  &  =k_{0}\ l_{[\alpha}n_{\beta]}+k_{1}\ m_{[\alpha}\bar
{m}_{\beta]}\\
A_{\alpha\beta;\nu}  &  =0
\end{align}
$A_{\alpha\beta}$ is therefore a KY solution. Note that $l_{[\alpha}n_{\beta
]}\sim dt\wedge dz$ and $m_{[\alpha}\bar{m}_{\beta]}\sim dx\wedge dy$. These
are the volume elements of the BR manifold.

\end{document}